\documentclass[11pt]{article}

\usepackage[utf8]{inputenc}
\usepackage[T1]{fontenc}
\usepackage{lmodern}
\usepackage[margin=1in]{geometry}
\usepackage{microtype}
\usepackage{graphicx}
\usepackage{booktabs}
\usepackage{amsmath}
\usepackage{textcomp}         
\usepackage{xcolor}
\usepackage{enumitem}
\usepackage{caption}
\usepackage{listings}
\usepackage[hidelinks]{hyperref}

\newcommand{\doi}[1]{\href{https://doi.org/#1}{doi:#1}}

\definecolor{codebg}{gray}{0.97}
\definecolor{codecomment}{rgb}{0.35,0.45,0.35}
\definecolor{codekw}{rgb}{0.10,0.20,0.55}
\lstdefinelanguage{ZetaSQL}{
  morekeywords={CREATE,TABLE,INTEGER,PRIMARY,KEY,TEXT,INSERT,INTO,VALUES,BRANCH,
    SET,SELECT,FROM,RESET,MERGE,DROP,ALTER,REBASE},
  sensitive=false,
  morecomment=[l]{--},
  morestring=[b]',
}
\title{\vspace{-1.5em}\bfseries Zeta-Lite: A Concurrent, Branchable In-Browser SQL
Database for Agentic Memory}
\author{Gene Zhang\thanks{Correspondence: \texttt{genegzhang@gmail.com}.}\\[0.4em]
\normalsize System \& code: \url{https://github.com/genezhang/zeta-lite}}
\date{August 2026 \\[0.3em]\normalsize v0.1 preview}

\begin{document}
\maketitle

\begin{abstract}
\noindent
The browser has become a first-class database host: applications increasingly
want to store, query, and reason over structured data entirely on the client ---
for privacy, offline operation, local-first collaboration, and, most recently, as
durable memory for in-browser AI agents. The prevailing way to get SQL in the
browser, compiling PostgreSQL to WebAssembly (PGlite), inherits PostgreSQL's
process model: a single backend connection that executes one statement at a time
and blocks. That model cannot express concurrent transactions, and it leaves
richer capabilities --- semantic search, graph queries, database branching --- to
whatever the compiled server happens to include.

We present \textbf{zeta-lite}, the browser form factor of the Zeta database
engine: a WebAssembly build that compiles the \emph{same} parser, planner,
optimizer, and executor as the Zeta server down to a \textbf{2.87\,MB gzipped}
artifact. Zeta-lite keeps the engine's log-centric asynchronous MVCC core, and
this single design choice yields two capabilities no other in-browser SQL engine
provides. First, \textbf{overlapping snapshot-isolated transactions on a single
thread}: multiple transactions hold distinct read/commit timestamps and
interleave, with snapshot-isolation conflict detection between them, without
threads or \texttt{SharedArrayBuffer}. Second, \textbf{copy-on-write database
branching} --- whole-database fork, merge, and rebase --- which falls out of the
MVCC log almost for free and is, to our knowledge, absent from every other browser
SQL engine and rare even in servers. On top of these, zeta-lite exposes a
feature-complete PostgreSQL surface (joins, CTEs, window functions, JSONB with GIN
indexes, full-text search, HNSW vector search, SQL/PGQ graph queries,
multi-database) and snapshot-to-OPFS durability that needs no worker, no
\texttt{SharedArrayBuffer}, and no cross-origin isolation headers. Across Chrome,
Firefox, and a native reference runtime, zeta-lite sustains 268k--315k point
reads/s and holds a mixed read/write workload flat over millions of operations,
with in-browser throughput within $\sim$5--15\% of native. We argue that a
completeness/concurrency/size trade-off usually taken for granted is not
fundamental: the log-centric MVCC design collapses it, and the result is a small,
fully-featured, concurrent SQL database that is an especially good fit for agentic
memory --- where cheap branchable state lets an agent explore, inspect, and commit
or discard speculative work. This is a v0.1 preview.
\end{abstract}

\vspace{0.5em}
\noindent\textbf{Keywords:} in-browser database; WebAssembly; MVCC; snapshot
isolation; database branching; agentic memory; OPFS; local-first.

\vspace{0.25em}
\noindent\textbf{ACM CCS:} Information systems $\rightarrow$ Database management
system engines; Software and its engineering $\rightarrow$ Runtime environments;
Computing methodologies $\rightarrow$ Intelligent agents.

\paragraph{Availability.}
The compiled zeta-lite engine is published to npm as \texttt{zeta-lite} and
attached to GitHub Releases; it is free for any use, including commercial, though
the engine source is closed (available separately under NDA). The hand-authored
surface --- the interactive SQL playground, the SQL reference, and the full
evaluation apparatus used in this paper --- is public in the repository above.
Every measurement in \S7 is reproducible from it against the \emph{published}
artifact, with no engine-source access required: the throughput and
snapshot-isolation benchmarks (\texttt{playground/bench.mjs},
\texttt{playground/bench.html}), the in-browser soak harness
(\texttt{playground/endurance.html}), the recorded native and two-browser soak
reports (\texttt{docs/benchmarks/}), and the methodology
(\texttt{docs/benchmarks/README.md}).

\section{Introduction}

Data has been moving toward the client. Applications that once round-tripped every
query to a server increasingly want to hold structured data in the browser itself:
local-first apps keep working offline and sync opportunistically; privacy-sensitive
tools keep user data on the device by construction; collaborative editors resolve
state locally for latency. The most recent driver is AI agents that run in the
browser and need somewhere durable to keep what they learn --- a memory that is
queryable, structured, and private to the user's device. In all of these, the
database is no longer a remote service but a component that ships with the page.

The dominant way to get real SQL in the browser today is to compile PostgreSQL to
WebAssembly, as PGlite does. This is a genuine achievement and it established that
the demand is real. But it inherits PostgreSQL's process architecture: a single
backend connection that runs one statement at a time and blocks until it finishes.
That architecture cannot represent two transactions in flight at once, so the
concurrency semantics an application can rely on are limited to what a single
serial connection provides. It also fixes the feature set to whatever the compiled
server includes, with capabilities like semantic search, graph queries, or
database branching either absent or bolted on separately.

We take a different architectural bet. \textbf{Zeta-lite} is the browser form
factor of the Zeta database engine, a single Rust codebase that also builds as an
embedded library, an OLTP/HTAP server, a cluster, and an OLAP engine. The browser
build compiles the same parser, planner, optimizer, and executor to a 2.87\,MB
gzipped WebAssembly artifact and keeps the engine's defining feature --- a
\textbf{log-centric asynchronous MVCC core}. From that one design choice, two
capabilities follow that no other in-browser SQL engine provides, and both are
consequences of the same mechanism rather than separately engineered features. The
first is \textbf{overlapping snapshot-isolated transactions on a single thread}:
because a transaction's view is a timestamp over an append-only log rather than
ownership of a connection, many transactions can be open at once, each on its own
snapshot, interleaving at statement boundaries with snapshot-isolation conflict
detection --- with no threads and no \texttt{SharedArrayBuffer}. The second is
\textbf{copy-on-write database branching}: whole-database fork, merge, and rebase,
which the log makes nearly free because the row versioning a branch needs already
exists for isolation.

On top of these, zeta-lite carries a feature-complete PostgreSQL surface --- joins,
CTEs, window functions, JSONB with GIN indexes, full-text search, HNSW vector
search, SQL/PGQ graph queries, multi-database --- and persists via a snapshot to
the Origin Private File System that needs no worker, no \texttt{SharedArrayBuffer},
and no cross-origin-isolation headers.

\paragraph{Contributions.}
\begin{enumerate}[leftmargin=1.5em,itemsep=0.25em]
\item \textbf{Overlapping snapshot-isolated transactions on a single wasm thread},
with an explicit account of the boundary (transaction-lifetime overlap and
cooperative interleaving between statements, not sub-statement parallelism).
\item \textbf{A feature-complete PostgreSQL surface in a 2.87\,MB gzipped
artifact} --- in the same size class as the single-connection PGlite baseline
while carrying a strictly larger feature set.
\item \textbf{Copy-on-write database branching in the browser} --- whole-database
fork / merge / rebase, rare in any SQL engine and (to our knowledge) absent from
every other in-browser one --- obtained almost for free from the MVCC log, and its
use as speculative exploration state for agents.
\item \textbf{Snapshot-to-OPFS durability} without workers,
\texttt{SharedArrayBuffer}, or COOP/COEP headers.
\item \textbf{A design-space account of the browser host interface} --- why
zeta-lite binds to Web platform APIs through JavaScript rather than using WASI, and
when WASI would apply instead.
\item \textbf{The Zeta-family position}: one engine codebase specialized by build
target, with the browser as its smallest and most demanding form factor.
\end{enumerate}

We argue throughout that the completeness/concurrency/size trade-off usually taken
for granted --- small browser engines drop features or concurrency; full engines
are large or need threads --- is not fundamental. The log-centric MVCC design
collapses it, and the resulting artifact is an especially good fit for agentic
memory, where cheap branchable state lets an agent explore, inspect, and then
commit or discard speculative work. Zeta-lite is a v0.1 preview; \S8 states its
boundaries plainly.

\section{Background and Related Work}

\paragraph{SQL in the browser.}
PGlite~\cite{pglite} compiles PostgreSQL to WebAssembly and is the closest point of
comparison: it offers a large, faithful PostgreSQL surface and OPFS persistence,
and it demonstrated the appetite for client-side SQL. Its constraint is
architectural rather than incidental --- it runs PostgreSQL in single-user mode, a
single connection whose internal mutexes admit one transaction at a
time~\cite{pgliteapi}, and its v0.4 connection multiplexer serializes multiple
clients through that one engine rather than running their transactions
concurrently~\cite{pglitev04}. This serial-transaction model is exactly the axis on
which zeta-lite differs (\S7.1). The other major family is SQLite in WebAssembly,
the most widely deployed in-browser SQL engine: the official SQLite-wasm
build~\cite{sqlitewasm} and wa-sqlite pair the engine with an OPFS virtual file
system, and at roughly 400\,KB gzipped~\cite{kanopy} it is far smaller than either
PGlite or zeta-lite --- a genuine advantage where footprint dominates and the SQL
surface can be narrow. Its trade-offs run the other way on the axes this paper
cares about. Its concurrency model is single-writer with concurrent readers (WAL),
so it too has no notion of two write transactions overlapping under snapshot
isolation. Its durable-persistence path typically uses a synchronous access handle
(\texttt{createSyncAccessHandle}), which is only available inside a Web Worker and,
with shared memory, pulls in the cross-origin-isolation (COOP/COEP) headers that
zeta-lite's snapshot-to-OPFS model avoids entirely (\S5.2). And it is SQLite, not
PostgreSQL --- no JSONB/GIN, HNSW vector search, SQL/PGQ, or database branching in
the engine itself. The three engines thus occupy distinct points: SQLite-wasm is
small and narrow; PGlite matches zeta-lite's size with a faithful but
serial-transaction Postgres; zeta-lite matches PGlite's size while adding
overlapping SI and branching. DuckDB-wasm~\cite{kanopy} targets analytical,
columnar workloads in the browser (roughly 2.8\,MB gzipped) and is a different
point in the design space --- OLAP rather than transactional OLTP --- which the
Zeta family addresses with a separate, larger form factor rather than the browser
build. We mention it for completeness of the landscape, not as a transactional
comparison.

\paragraph{MVCC and snapshot isolation.}
Zeta-lite's transactional core is a multiversion concurrency-control engine
providing snapshot isolation, the isolation level characterized by Berenson et
al.~\cite{berenson}, and related to PostgreSQL's own SI and serializable-snapshot-
isolation implementations. We claim no novelty in the isolation level itself; the
contribution is delivering \emph{overlapping} SI transactions in a single-threaded
browser engine at a small artifact size, which the log-centric formulation makes
possible without threads.

\paragraph{Database branching.}
Branching of database state is a capability that a handful of systems have made
central. Dolt~\cite{dolt} implements Git-style versioning --- branches, diffs, and
merges --- in a purpose-built storage engine. Neon~\cite{neon} provides
copy-on-write branches for PostgreSQL at the storage layer, so a branch is a cheap
fork of the page store. PlanetScale~\cite{planetscale} exposes branching as a
schema-change and deploy workflow over MySQL. What these share is that branching is
a server- or storage-tier feature, often the product's headline, built as dedicated
machinery. Zeta-lite differs on two counts: its branching is a direct consequence
of the MVCC log rather than a bespoke subsystem (\S3.6), and it runs entirely in
the browser, which none of these do.

\paragraph{wasm as an engine host.}
Compiling data-system engines to WebAssembly is an active area, spanning both the
browser (via JavaScript bindings) and server-side wasm runtimes (via WASI). \S4
addresses the host-interface choice directly, since it is a recurring source of
confusion and a deliberate design point for zeta-lite.

\paragraph{The delta.}
To our knowledge, no in-browser SQL engine offers overlapping snapshot-isolated
transactions, and none offers whole-database branching. Zeta-lite provides both, at
a gzip size in the same class as the single-connection PGlite baseline, as
consequences of one architectural choice.

\section{System Architecture}

\begin{figure}[htbp]
  \centering
  \includegraphics[width=\linewidth]{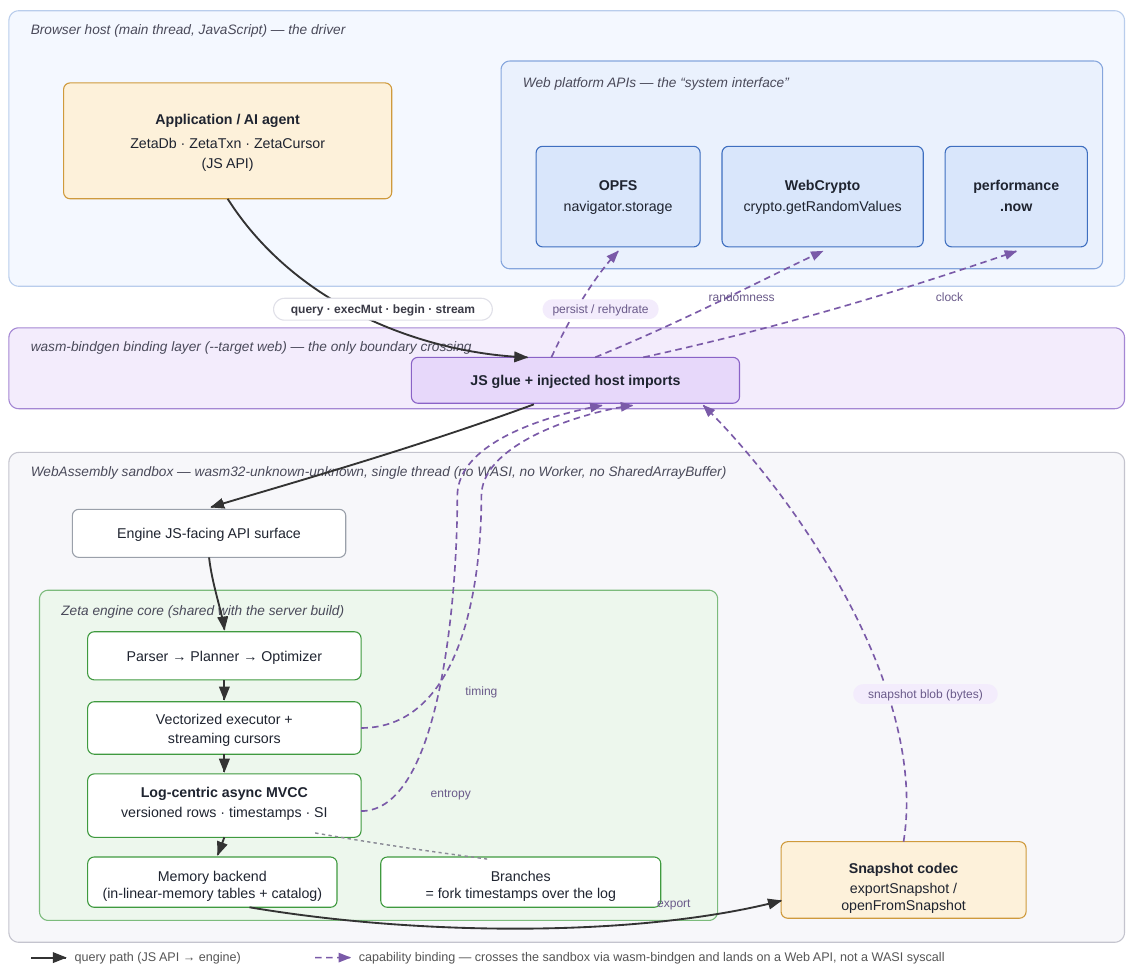}
  \caption{Zeta-lite component architecture: engine in the wasm sandbox,
  capabilities bound to Web APIs (not WASI). Solid edges are the query path (JS API
  $\rightarrow$ engine); dashed edges are capability bindings that cross the sandbox
  via \texttt{wasm-bindgen} and land on a Web API rather than a WASI syscall.}
  \label{fig:arch}
\end{figure}

Figure~\ref{fig:arch} gives the whole picture: the engine core executes inside the
WebAssembly sandbox, and every capability it needs from the outside world is
supplied by the browser through a \texttt{wasm-bindgen} binding layer --- \emph{not}
by WASI. The two OS-like capabilities a database normally requires, persistence and
randomness, land on Web platform APIs (OPFS and WebCrypto); \S4 develops why this,
and not a WASI syscall surface, is the correct host interface for the browser. The
subsections below walk the components from the top of the figure down.

The browser host is the driver: the application calls down into the engine (solid
edges, the query path), and the capability bindings cross back out (dashed edges) to
the host's own Web APIs. Each dashed edge terminates on a Web API, the concrete form
of ``the browser is the system interface'' (\S4.2). Notably absent is any
WASI/\texttt{wasip1} layer, any Web Worker, and any \texttt{SharedArrayBuffer}:
persistence is a snapshot blob handed to OPFS on the main thread (\S5), so the
design needs no cross-origin-isolation headers.

\subsection{The Zeta family and the compile-down bet}

Zeta-lite is not a separate engine. It is the smallest point in a family of form
factors --- embedded, OLTP/HTAP server, cluster, and OLAP/warehouse --- that share
one Rust codebase and are distinguished by Cargo feature selection at build time.
The browser build is \texttt{cargo build -{}-no-default-features -{}-features wasm},
which selects the in-memory (``Memory'') storage backend and compiles out the
pgwire protocol layer, the on-disk page store, and the server runtime, while
retaining the full SQL front end and execution engine. The bet the family makes is
that the \emph{hard} part of a database --- a correct parser, a cost-based planner,
an MVCC executor with snapshot isolation --- is worth writing once and specializing
by target, rather than reimplementing per form factor. Zeta-lite is the strongest
test of that bet: if the same engine can be made to fit a browser tab at under
3\,MB while keeping server-grade semantics, the shared-core design pays off at both
ends of the size spectrum.

The practical consequence for this paper is that zeta-lite's SQL surface and its
transactional semantics are not browser-specific reimplementations; they are the
server's, minus the features the Memory backend cannot support (durable page-level
persistence; see \S5). When we claim ``the same SQL as the Zeta server,'' we mean
the same code path.

\subsection{The log-centric asynchronous MVCC core}

The engine is organized around a \textbf{log as the source of truth}. Writes append
versioned records; each row version carries the timestamp of the transaction that
produced it. A transaction acquires a \textbf{read timestamp} when it begins and
sees exactly the versions committed at or before that timestamp --- its consistent
snapshot --- and acquires a \textbf{commit timestamp} when it commits. Snapshot
isolation follows directly: a transaction reads its snapshot regardless of
concurrent commits, and at commit the engine checks for write-write conflicts
against versions committed since the transaction's read timestamp, aborting if
another transaction has written a row it also wrote.

This is a standard MVCC formulation; what matters for the browser is what it makes
possible without threads. Because a transaction's view is defined by a timestamp
over an append-only log rather than by exclusive ownership of a connection,
\textbf{multiple transactions can be open at once}, each on its own snapshot, and
interleave at statement boundaries on a single thread. Transaction A can hold its
snapshot open across several statements while transaction B commits; A continues to
see its original snapshot, and if A later writes a row B has since committed, A's
commit is the one that aborts. \S7.1 demonstrates both halves of this behavior
directly.

\subsection{The query path}

Statements flow through the same stages as in the server build: a
PostgreSQL-dialect parser produces an AST; a planner and cost-based optimizer
produce a physical plan (join ordering, index selection, predicate pushdown); and a
vectorized-batch executor pulls rows. The reachable surface includes inner and
outer joins, common table expressions, correlated and uncorrelated subqueries,
window functions, aggregates with \texttt{GROUP BY}/\texttt{HAVING},
\texttt{INSERT}/\texttt{UPDATE}/\texttt{DELETE} with \texttt{RETURNING}, and
secondary indexes including GIN over JSONB and HNSW over vector columns. ``Same SQL
as the server'' is meant literally at the code level --- the JS-facing API
(\texttt{query}, \texttt{execMut}, \texttt{execDdl}, \texttt{stream},
\texttt{begin}) is a thin wasm-bindgen wrapper over the identical planner and
executor.

\subsection{Streaming cursors}

Query results are available either fully materialized (\texttt{query} returns
\texttt{\{ columns, rows \}}) or through a \textbf{streaming cursor}
(\texttt{stream} returns a handle whose \texttt{next()} yields one row at a time).
The cursor pulls rows lazily in bounded batches, so peak memory is $O(\text{batch})$
rather than $O(\text{result-size})$. This is not a convenience feature in the
browser: wasm linear memory is a single growable \texttt{ArrayBuffer} that never
shrinks back to the host, so a query that materializes a large result permanently
inflates the tab's footprint. Streaming keeps a scan over a large table bounded,
which is the difference between a workable and an unworkable memory profile for an
agent iterating over its stored history.

\subsection{The Memory backend}

The storage substrate for the browser build is a set of in-memory, MVCC-versioned
tables. There is no page store and no filesystem dependency; the entire database ---
catalog and row versions --- lives in wasm linear memory. Durability is achieved by
serializing that state to a byte blob and persisting it out-of-band (\S5), not by
writing pages on each commit. This is the deliberate boundary of the smallest form
factor: it trades durable-by-default persistence for a tiny, dependency-free
artifact that runs in any browser context.

\subsection{Database branching as a consequence of the log}

The most distinctive capability zeta-lite exposes is \textbf{copy-on-write branching
of the whole database}, and its cost is what makes it notable: because the engine
already versions every row by timestamp for snapshot isolation, a branch needs no
separate storage machinery. \textbf{A branch is a fork timestamp
(\texttt{fork\_ts}) over the shared log.} \texttt{CREATE BRANCH feat} records the
current timestamp as the branch's fork point; the branch thereafter sees
\emph{main-as-of-\texttt{fork\_ts} plus its own writes}, and main does not see the
branch's writes at all. \texttt{MERGE BRANCH feat} publishes the branch's delta ---
data \emph{and} catalog changes --- back into main and retires the branch;
\texttt{ALTER BRANCH feat REBASE} moves the fork point forward to the latest main,
re-anchoring a long-lived branch on newer committed state; \texttt{DROP BRANCH feat}
discards it. Branch selection is per handle (\texttt{setBranch}), re-resolved on
each statement, so a dropped or rebased branch surfaces a clear error on its next
use rather than silently reading stale state.

The point is that this is the \emph{same insight} that makes \S7.1's transaction
overlap cheap, applied at a coarser grain: SI overlaps transactions by giving each a
timestamp view of the log; branching overlaps entire lines of history the same way.
Whole-database branching is rare in any SQL engine --- PostgreSQL has none natively;
Neon and Dolt build entire products around branching at the storage and server
tiers --- and, as far as we are aware, no other in-browser SQL engine offers it at
all. Zeta-lite gets it at 2.87\,MB because the versioning it requires was already
there for correctness. One honest edge remains in v0.1: \texttt{exportSnapshot()}
does not yet serialize branches, so a database must merge or drop its branches
before it is persisted; we treat this as a limitation (\S8), not a design boundary.

\S6.2 develops why this capability is a particularly natural fit for AI agents.

\section{Executing in the Browser: wasm, not WASI}

A question that recurs whenever an engine is compiled to WebAssembly is whether it
uses WASI, the WebAssembly System Interface. Zeta-lite deliberately does not, and
the reason is not an omission but a correct reading of what the browser is. This
section makes the distinction precise, because conflating the two layers is the most
common framing error in browser-database work.

\subsection{Two different layers}

\textbf{WebAssembly} is a portable bytecode and a sandboxed linear-memory virtual
machine. By itself it has no ambient authority whatsoever --- no files, no clock, no
network, no randomness, no standard I/O. A wasm module can compute over its own
linear memory and call exactly the functions its host explicitly provides as
imports, and nothing else. That sandbox is the security model, not a limitation to
be worked around.

\textbf{WASI} is a standardized \emph{set of host imports} --- effectively a syscall
ABI --- that grants a wasm module POSIX-like operating-system capabilities:
\texttt{fd\_read}/\texttt{fd\_write} for file descriptors, \texttt{path\_open} for
the filesystem, \texttt{clock\_time\_get} for time, \texttt{random\_get} for
entropy. WASI exists so that a wasm module can run \emph{outside the browser} --- in
a server runtime such as Wasmtime or WasmEdge, or under Node's WASI shim --- and
still open files and read the clock against a capability-scoped host. In one
sentence: WASI is wasm's operating-system interface for \textbf{non-browser} hosts.

\subsection{Why the browser needs no WASI}

The browser is not a WASI host, and never was. In the browser, the ``system
interface'' is already present in a richer form: it is JavaScript together with the
Web platform APIs. A wasm module reaches persistence through the Origin Private File
System, entropy through \texttt{crypto.getRandomValues}, time through
\texttt{performance.now}, and everything else through JavaScript bindings its host
injects as imports. There is no need for WASI's syscall ABI because the capabilities
WASI standardizes for server runtimes are supplied, in the browser, by the Web
platform through a different and already-universal interface.

\begin{table}[htbp]
\centering
\caption{The two OS-like capabilities a database needs, as WASI syscalls (server
runtimes) versus Web/JS bindings (zeta-lite in the browser).}
\label{tab:wasi}
\small
\begin{tabular}{@{}p{0.16\linewidth}p{0.36\linewidth}p{0.40\linewidth}@{}}
\toprule
\textbf{Capability} & \textbf{WASI syscall (non-browser)} & \textbf{Web/JS binding (zeta-lite, browser)} \\
\midrule
Persistence & \texttt{path\_open} / \texttt{fd\_write} / \texttt{fd\_read} & OPFS via \texttt{navigator.storage.\allowbreak getDirectory} \\
Randomness & \texttt{random\_get} & \texttt{crypto.getRandomValues} (\texttt{getrandom} \texttt{wasm\_js} backend) \\
Time & \texttt{clock\_time\_get} & \texttt{performance.now} / \texttt{Date.now} \\
Console / logging & \texttt{fd\_write} to stdout/stderr & \texttt{console.*} via bindings \\
\bottomrule
\end{tabular}
\end{table}

These are exactly the dashed capability edges of Figure~\ref{fig:arch}: each row is
a binding that crosses the sandbox boundary and terminates on a Web API rather than
a WASI syscall.

\subsection{Zeta-lite's binding strategy}

Concretely, zeta-lite compiles to the \textbf{\texttt{wasm32-unknown-unknown}}
target --- the ``unknown host'' target that assumes no operating system and no WASI
--- and then runs \texttt{wasm-bindgen -{}-target web} to generate the JavaScript
shim that supplies every host capability the engine needs as an import. The two
capabilities a database normally takes from the OS are provided the browser-native
way. Randomness is routed through Rust's \texttt{getrandom} crate configured with
\texttt{-{}-cfg getrandom\_backend="wasm\_js"}, which calls
\texttt{crypto.getRandomValues} rather than WASI's \texttt{random\_get}. Persistence
is the OPFS transport described in \S5, built on \texttt{navigator.storage}, rather
than WASI's \texttt{path\_open}/\texttt{fd\_write} against a preopened directory. The
engine core is unaware of either choice; it calls a small platform trait whose
browser implementation is these bindings.

\subsection{When WASI would matter, and the trajectory}

WASI is not irrelevant to the Zeta family --- it is simply the wrong interface for
\emph{this} form factor. A wasm build intended to run server-side or as a CLI under
a wasm runtime would target \texttt{wasm32-wasip1} and use WASI to reach a real
filesystem, which is a different deployment than the in-browser one this paper
concerns. It is also worth noting the direction of travel: the WASI Preview 3 /
Component Model effort and interfaces such as \texttt{wasi-filesystem} are
converging wasm toward portable, capability-scoped host APIs that could eventually
be provided in-browser too. Today, however, the browser story is JavaScript
bindings, and that is the mainstream and correct choice for a browser database.
Zeta-lite is not missing a feature by omitting WASI; it is using the host interface
its target actually provides.

\section{Persistence: Snapshot-to-OPFS Durability}

The Memory backend (\S3.5) keeps the whole database in wasm linear memory, so
durability is a separate, deliberate mechanism rather than a property of each write.

\subsection{The model}

Durability in zeta-lite is \textbf{whole-database snapshotting}.
\texttt{exportSnapshot()} serializes the entire state --- catalog, row versions, and
the timestamp high-water mark --- into a single \texttt{Uint8Array}; the application
writes that blob to the Origin Private File System, and on the next load
\texttt{openFromSnapshot(bytes)} rehydrates a fresh database from it. The round-trip
is exact: an exported-then-restored database is indistinguishable from the original,
which \S7.5's validation covers, including a database carrying a vector column. The
snapshot is the unit of durability, and it is explicit --- the application decides
when to checkpoint.

\subsection{Why snapshot, not sync-per-write}

The alternative --- a storage engine that fsyncs each commit to the file system ---
is what SQLite-wasm's OPFS VFS approximates, and it comes at a specific cost in the
browser. Synchronous OPFS access (\texttt{createSyncAccessHandle}), which a
sync-every-write engine needs, is only available inside a Web Worker, and driving a
database from a worker with shared memory pulls in \texttt{SharedArrayBuffer}, which
in turn requires the page to be cross-origin isolated via COOP and COEP response
headers. Those headers are a real deployment burden: they change how the page may
embed and be embedded, and they are not always within the developer's control.

Snapshot-on-demand sidesteps all of it. Because a checkpoint writes one whole blob
at a chosen moment rather than syncing on every commit, it can use the \textbf{plain
asynchronous OPFS API} (\texttt{getFileHandle} + \texttt{createWritable}) directly
on the main thread. Zeta-lite therefore runs in any browser context that has OPFS,
with no worker, no \texttt{SharedArrayBuffer}, and no special headers. This is a
genuine simplification the in-memory design buys: the persistence path is a thin
async transport, and the engine core never blocks on I/O. Snapshot export is also
cheap in practice --- \S7.3 measured export$\rightarrow$rehydrate cycles averaging
44\,ms for blobs up to $\sim$2\,MB.

\subsection{The durability window}

The honest cost of this model is that durability is coarse-grained. A committed
transaction is durable only \textbf{as of the last snapshot the application
persisted to OPFS}; anything committed after that snapshot is lost on a crash,
reload, or tab close. This is snapshot durability, not per-commit durability, and it
places a requirement on the application: checkpoint after writes that must survive.
For the target workloads --- local-first state and agentic memory, where a
checkpoint after a meaningful unit of work is natural --- this is an acceptable and
explicit contract rather than a hidden hazard. We note it as a first-class part of
the persistence design, and \S8 records the related v0.1 limitation that snapshots
do not yet capture branches.

\section{The SQL Surface as Agentic Storage}

The preceding sections argue the engine on systems grounds. This section turns to
the use case that motivates the title: zeta-lite as durable memory for AI agents
running in the browser. The claim here is subordinate to the architecture --- we do
not evaluate an agent task (\S7 measures the engine, not a workload on top of it)
--- but the fit is worth making concrete, because the capabilities an agent's memory
wants line up unusually well with what a full SQL surface already provides, and
because one of them, branching, is the capability agents most need and can least
often get.

\subsection{The surface an agent wants}

An agent's memory is heterogeneous, and each part maps to something the engine
already has. \textbf{JSONB with GIN indexes} stores the semi-structured output of
tool calls and API responses without a rigid schema, and queries it by containment.
\textbf{Full-text search} covers keyword retrieval over stored text. \textbf{HNSW
vector search}, wired to an embedder through \texttt{embed()}, provides semantic
recall --- nearest-neighbor over embeddings the agent has stored. \textbf{SQL/PGQ
graph queries} let the agent traverse relationships in its own data --- entities,
references, derivations --- as a graph rather than through hand-written recursive
joins. \textbf{Multi-database} gives per-task or per-session logical namespaces over
one shared catalog. The point is not that any one of these is unique to zeta-lite,
but that an agent memory usually assembled from several specialized stores --- a
vector index here, a document store there, a graph database beside them --- is here
a single 2.87\,MB SQL engine with transactional consistency across all of it. On the
embedding boundary: \texttt{embed()} ships no bundled model, so an application
registers a synchronous embedder from JavaScript via \texttt{setEmbedFn}, or
computes vectors in JS and binds them; the callback must be synchronous because
query execution is synchronous on the single thread (\S4.3).

\subsection{Database branching as agent exploration state}

Of the capabilities in \S6.1, database branching is the one an agent loop most
naturally wants and the one it can least often get. The right way to see it is as
\textbf{speculative execution over persistent state}. An agent frequently needs to
try something whose outcome it cannot predict: apply a tentative plan, run a
sequence of tool calls that mutate stored data, or test a hypothesis against its
memory --- and then keep the result only if it worked. Without branching, an agent's
options are all poor: snapshot the entire database and restore on failure (coarse
and expensive as memory grows), copy the working set into a scratch table and
reconcile by hand (error-prone, and it does not cover schema changes), or maintain
an application-level undo log (a database inside the database).

Branching turns this into a first-class operation. The agent forks a branch, does
its speculative work in isolation from the durable line, inspects the outcome, and
then either \textbf{merges} --- publishing the delta, schema changes included ---
or \textbf{drops} it, reverting to exactly the pre-fork state at no cleanup cost. A
long-running exploration that must stay current with facts the agent has learned in
the meantime can \textbf{rebase} its fork point onto the latest main rather than
starting over. Because a branch is a fork timestamp over the MVCC log (\S3.6), this
is cheap enough to do per hypothesis, it is whole-database (both rows and catalog
fork, so an agent may create tables on a branch), and it is entirely client-side ---
no server round-trip stands between a thought and the state that explores it.

A minimal illustration, mirroring the shipped branching example:

\begin{lstlisting}
CREATE TABLE t (id INTEGER PRIMARY KEY, v TEXT);
INSERT INTO t VALUES (1, 'on main');
CREATE BRANCH feat;
SET zeta_branch = 'feat';
INSERT INTO t VALUES (2, 'only on feat');   -- isolated on the branch
SELECT * FROM t;                            -- branch sees id 1, 2
RESET zeta_branch;
SELECT * FROM t;                            -- main still sees only id 1
MERGE BRANCH feat;
SELECT * FROM t;                            -- main now sees id 1, 2
\end{lstlisting}

This is the substrate the companion system, \textbf{zengram-lite} --- an agentic
memory system for in-browser agents [forthcoming] --- is built on: vector-indexed
semantic memory, graph queries over the agent's own relationships, and branchable
state for exploration, all in the same 2.87\,MB engine.

\section{Evaluation: Throughput under Concurrency, Coverage, and Size}

We evaluate four things: that the concurrency claim is real (\S7.1), that
single-thread throughput is more than adequate and the browser tax is small
(\S7.2), that the engine is stable under sustained load (\S7.3), and that all of
this fits the size and coverage envelope claimed (\S7.4--\S7.5). All measurements
run against the \textbf{published} wasm artifact, so they are reproducible from the
public repository without engine-source access; the two throughput/concurrency
benches and the recorded soak-test output are in
\texttt{docs/benchmarks/}~\cite{artifacts}.

\paragraph{Setup.}
One machine (AMD Ryzen AI MAX+ 395, 32 threads, of which the engine uses one),
artifact \texttt{zeta\_wasm\_bg.wasm} at ${\sim}9.7$\,MB raw / \textbf{2.87\,MB
gzipped}. Browser targets: Chrome 152 (V8) and Firefox 154 (SpiderMonkey); native
reference: bun 1.3.14. The engine is single-threaded throughout ---
\texttt{crossOriginIsolated} is \texttt{false} and no \texttt{SharedArrayBuffer} is
used, which the bench prints to make the absence of thread parallelism auditable.
One ``op'' is one API call: a single \texttt{execMut} (one autocommitted
INSERT/UPDATE/DELETE), one \texttt{query} (a point SELECT by primary key), or one
\texttt{begin}/\texttt{execMut}/\texttt{commit} transaction, as labeled. These are
single-row OLTP units, not batches. Figures are representative single runs, not
averaged; they reproduce in shape, not last digit.

\subsection{Snapshot isolation under contention}

This is the result the architecture exists to produce. Two transactions are opened
on overlapping snapshots and both \texttt{UPDATE} the same row; under snapshot
isolation exactly one may commit and the other must abort with a write-write
conflict. Repeated 5,000 times, the outcome is identical across all three runtimes:

\begin{table}[htbp]
\centering
\caption{Snapshot-isolation conflict detection under contention. Two transactions
on overlapping snapshots update the same row; exactly one commits.}
\label{tab:si}
\begin{tabular}{lrrr}
\toprule
 & \textbf{Chrome 152} & \textbf{Firefox 154} & \textbf{bun 1.3.14} \\
\midrule
Rounds (A and B both update row 1) & 5,000 & 5,000 & 5,000 \\
A commits & 5,000 & 5,000 & 5,000 \\
B commits & 0 & 0 & 0 \\
\textbf{B aborted (write-write conflict)} & \textbf{5,000 / 5,000} & \textbf{5,000 / 5,000} & \textbf{5,000 / 5,000} \\
\bottomrule
\end{tabular}
\end{table}

The complementary case confirms that overlap itself is permitted, not merely
tolerated: eight transactions opened concurrently on distinct snapshots, each
writing a \textbf{disjoint} key, all commit --- 40,000 commits, zero conflicts.
Overlap is allowed; only genuine write-write conflicts abort. The significance is
not the abort rate but that the workload \emph{exists}: PGlite runs PostgreSQL in
single-user mode --- one connection, with an internal transaction mutex that admits
only one transaction at a time --- so it cannot open a second concurrent transaction
to conflict with in the first place. (PGlite v0.4's connection multiplexer lets
multiple clients attach, but it serializes their work through the single engine
rather than executing transactions concurrently, so the semantic is unchanged.) This
table therefore has no single-connection baseline to compare against --- which is
the point.

\subsection{Throughput and the browser tax}

Each row below is 50,000 ops. The headline is twofold: single-thread throughput is
high in absolute terms, and \textbf{the in-browser numbers are within
$\sim$5--15\% of native bun}, with the two browser engines agreeing closely --- the
result is not an artifact of a particular JavaScript engine.

\begin{table}[htbp]
\centering
\caption{Throughput across two browsers and a native reference. Each row is 50,000
single-row OLTP operations.}
\label{tab:tput}
\begin{tabular}{lrrr}
\toprule
\textbf{Measurement} & \textbf{Chrome 152} & \textbf{Firefox 154} & \textbf{bun (native ref)} \\
\midrule
Serial INSERT (autocommit / op) & 121,743 ops/s & 98,619 ops/s & 116,474 ops/s \\
INSERT inside one txn (amortized) & 77,030 ops/s & 77,882 ops/s & 84,327 ops/s \\
Point SELECT (PK lookup) & 267,953 ops/s & 282,486 ops/s & 315,345 ops/s \\
Overlapping SI txns, K=8, disjoint & 60,286 ops/s & 58,824 ops/s & 62,568 ops/s \\
Sustained mixed R/W (10\,s) & 228,284 ops/s & 221,399 ops/s & 240,841 ops/s \\
\quad --- per-window drift over 10\,s & 2.2\% & 1.1\% & 3.8\% \\
\bottomrule
\end{tabular}
\end{table}

Per-operation latency, measured by the soak harness (\S7.3) at operation-class
granularity, is consistent with these rates: a point read has p50 4\,\textmu s / p99
8\,\textmu s, an insert p50 9\,\textmu s, an update p50 12\,\textmu s, and a full
begin/write/commit transaction p50 51\,\textmu s / p99 158\,\textmu s. A 4\,\textmu s
point read implies a single-thread read ceiling on the order of $10^5$--$10^6$
ops/s, which is what the 268k--315k point-SELECT throughput shows. We stress that
these are not ``fast database'' claims in an absolute, server-benchmark sense; they
establish that a single wasm thread has ample headroom over any realistic in-browser
workload \emph{while} providing isolation semantics no other browser engine offers.

\subsection{Sustained-load stability}

Throughput numbers say nothing about whether an engine degrades over time. We use a
soak harness that drives a mixed OLTP workload --- four seeded tables, a
periodically-refreshed long-lived reader transaction that pins a snapshot, periodic
\texttt{exportSnapshot}$\rightarrow$\texttt{openFromSnapshot} integrity cycles, and
branch create/merge/drop churn --- at a \textbf{deliberately modest, rate-limited
1,500 ops/s}. The rate limit is the point: the harness measures \emph{stability
under a realistic application load}, not peak throughput (that is \S7.2). We run it
two ways: natively (\texttt{crates/zeta-wasm/harness/endurance.mjs}), where an
unclamped timer gives trustworthy per-operation latencies, and \textbf{in the
browser} (\texttt{playground/endurance.html}), which measures the target environment
directly and exercises real OPFS persistence.

\paragraph{Native run (per-op latency, baseline stability).}
Over the native run: throughput stability is \textbf{1.00} (first-quarter 1,500
$\rightarrow$ last-quarter 1,500 ops/s, PASS floor 0.70); per-op-class latency is
low and is the reference cited in \S7.2 (point read p50 4\,\textmu s / p99
8\,\textmu s; range scan p50 46\,\textmu s; insert p50 9\,\textmu s; update p50
12\,\textmu s; a full begin/write/commit transaction p50 51\,\textmu s / p99
158\,\textmu s); and the run completes with zero SQL errors or invariant violations
across 4,250 branch cycles and eight snapshot round-trips (avg 44\,ms).

\paragraph{In-browser run (Chrome 152 and Firefox 154, 10-minute sustained phase,
825\,s total each).}
The soak holds up identically in the target environment, and the two browser engines
agree closely:

\begin{itemize}[leftmargin=1.5em,itemsep=0.35em]
\item \textbf{Throughput stability 1.00 in both browsers} --- first-quarter 1,502
$\rightarrow$ last-quarter 1,501 ops/s over \textbf{900,000 sustained operations},
in Chrome and Firefox alike. No drift.
\item \textbf{No leak, in both.} We measure the wasm linear-memory footprint
directly (\texttt{memory.buffer.\allowbreak byteLength}). During the insert-heavy sustained
phase it grows 131 $\rightarrow$ 195\,MB in each browser --- \emph{data-driven, not a
leak}: the workload only inserts, and wasm linear memory never shrinks back to the
host. The leak test is the \textbf{read-only settle phase}, over which the footprint
grows \textbf{+0\,MB} (32\,MB budget) in both --- flat, and in fact tighter than the
native run's +8\,MB.
\item \textbf{Real OPFS durability latency}, a measurement the native harness cannot
make, averaged over ten snapshot cycles per browser as the blob grew to
$\sim$4.8\,MB. A full round-trip --- \texttt{exportSnapshot} + OPFS write + OPFS read
+ rehydrate-and-verify --- is 38 + 48 + 5 + 85\,ms in Firefox and 34 + 19 + 2 +
78\,ms in Chrome. Persisting a multi-megabyte database to durable browser storage and
reading it back is a sub-200-ms operation in both; the one notable engine difference
is that Chrome's OPFS writes are roughly 2.5$\times$ faster than Firefox's (19\,ms vs
48\,ms average), a browser-implementation property rather than an engine one.
\item \textbf{Branch and snapshot churn} completed with \textbf{zero SQL errors or
invariant violations} --- 6,793 branch merge/drop cycles and 10 OPFS round-trips in
Firefox, 7,424 cycles and 10 round-trips in Chrome.
\end{itemize}

Two honesty notes. First, we report the sustained-phase memory growth (native
+460\,MB over its longer standard run; browser 131$\rightarrow$195\,MB) explicitly
rather than the flattering settle figure alone, because an insert-only workload
\emph{must} grow memory and the settle phase is the measurement that actually bears
on leaks. Second, \textbf{per-op latencies are taken only from the native run}:
browsers clamp \texttt{performance.now()} to $\sim$1\,ms resolution on pages that are
not cross-origin-isolated (a timing-side-channel mitigation), and zeta-lite is
deliberately not COI, so its in-browser sub-millisecond op timings are meaningless.
Every in-browser figure we do quote --- throughput over 5\,s windows, memory
footprint, and OPFS cost in tens of milliseconds --- is measured over intervals well
above that clamp.

\subsection{Artifact size}

The transfer size that matters is gzip: zeta-lite is \textbf{2.87\,MB gzipped}.
PGlite, the closest comparison, is officially described as ``under 3\,MB gzipped,''
with reported figures across versions and measurements ranging from roughly 2.6\,MB
to 3.3\,MB. Zeta-lite therefore sits in the same weight class as the
single-connection PGlite baseline --- neither is decisively smaller --- while
carrying a strictly larger set of capabilities: overlapping snapshot-isolated
transactions, HNSW vector search, SQL/PGQ graph queries, and whole-database
branching, none of which PGlite provides. The claim we make on size is not that
zeta-lite is the smallest browser SQL engine --- SQLite-wasm, at roughly 400\,KB
gzipped, is far smaller with a correspondingly narrower surface --- but that
\emph{at PGlite's size point}, a strictly more capable and more concurrent engine is
achievable. That is the concrete form of the paper's central argument: completeness,
concurrency, and size are not locked in a three-way trade-off at this point on the
curve.

\subsection{Functional coverage}

Surface completeness is verified by a validation harness that drives the real wasm
artifact through every shipped example and the concurrency demo --- \textbf{88
assertions across all examples} --- covering joins and aggregates, window functions,
JSONB with containment and GIN-backed predicates, full-text search, HNSW vector
nearest-neighbor, SQL/PGQ graph queries, multi-database resolution, the branching
workflow (on-branch / main-isolated / post-merge visibility), and snapshot
export/restore round-trips including a vector-column database. The harness runs
against the same \texttt{pkg-web/} build the playground loads, so ``the SQL the page
ships actually runs'' is a checked property, not a claim~\cite{validation}.

\section{Limitations and Future Work}

Zeta-lite is a v0.1 preview, and we state its boundaries plainly.

\paragraph{Concurrency is transaction-lifetime, not sub-statement.}
The overlap of \S7.1 is that multiple transactions stay open on distinct snapshots
and interleave \emph{between} statements; it is not simultaneous in-flight execution
of one query while another makes progress. The executor pulls rows synchronously, so
within a single statement there is no interior yield. True in-query parallelism
requires threads --- a future, header-gated build using \texttt{SharedArrayBuffer}
and wasm threads --- and is out of scope for the headers-free form factor described
here. We claim the overlap we deliver and no more.

\paragraph{Single-threaded.}
Following from the above, the engine uses one thread; the throughput ceilings of
\S7.2 are single-core. This is adequate for browser workloads by a wide margin
(\S7.3) but is a real ceiling for anything embarrassingly parallel.

\paragraph{Durability is snapshot-based.}
As \S5.3 details, a commit is durable only as of the last persisted snapshot. There
is no per-commit fsync in the browser build.

\paragraph{Snapshots do not yet capture branches.}
\texttt{exportSnapshot()} currently errors if branches exist; a database must merge
or drop its branches before it is persisted. This is an implementation gap, not a
design boundary --- unifying the snapshot codec with the branch representation is
planned --- but in v0.1 it means branch state and durable state are not
simultaneously available.

\paragraph{No OLAP/columnar engine.}
The browser build is a row-oriented OLTP engine. Analytical, columnar workloads are
addressed by a separate, larger form factor in the Zeta family, not by this
artifact.

\paragraph{Per-op latency is measured natively, not in-browser.}
As \S7.3 notes, the browser's $\sim$1\,ms \texttt{performance.now()} clamp on
non-cross-origin-isolated pages makes sub-millisecond in-browser op timings unusable,
so the per-operation latency table comes from the native harness. The in-browser
soak validates throughput stability, memory behavior, and OPFS cost directly (all
measured above the clamp), but fine-grained in-browser latencies would require a
cross-origin-isolated context we otherwise avoid.

\paragraph{Minor surface gaps.}
The playground's schema sidebar currently lists all databases' tables rather than
filtering per active database, and \texttt{embed()} is compiled in but ships no
bundled model --- an application registers a synchronous embedder from JavaScript
(\texttt{setEmbedFn}) or binds precomputed vectors. Neither affects the engine's SQL
semantics.

\paragraph{Future work}
centers on a threaded build for in-query parallelism, branch-aware snapshots, an
in-browser soak, and --- building on the branching and vector/graph surface --- the
companion agentic-memory system, zengram-lite.

\section{Conclusion}

The received wisdom is that an in-browser SQL database must choose: keep the artifact
small by dropping features and concurrency, or keep the features and accept size,
threads, and cross-origin-isolation headers. Zeta-lite is evidence that the choice
is false. By compiling the same log-centric MVCC engine as the Zeta server down to
\texttt{wasm32-unknown-unknown} and binding it to the Web platform through
JavaScript rather than WASI, it delivers overlapping snapshot-isolated transactions
and whole-database copy-on-write branching --- neither available in any other
in-browser SQL engine --- on top of a feature-complete PostgreSQL surface, in a
2.87\,MB gzipped artifact that needs no worker and no special headers. The two
headline capabilities are not two engineering efforts but one: both fall out of
treating a transaction's, and a branch's, view of the world as a timestamp over an
append-only log. That the same insight scales from interleaving transactions to
forking entire databases is the architectural point, and it is what lets a small
artifact behave like a much larger one.

The result is a particularly natural substrate for agentic memory. An agent that can
branch its database per hypothesis, search it semantically, query it as a graph, and
commit or discard speculative work at the cost of a timestamp has, in 2.87\,MB, a
memory system with properties that previously required a server. The companion system
built on exactly these primitives, zengram-lite, is the subject of ongoing work;
zeta-lite is the engine beneath it, and it stands on its own as the smallest and most
demanding point in the Zeta family.

\end{document}